\documentclass[aps,onecolumn,11pt,floatfix,altaffilletter,superscriptaddress,
               preprintnumbers,tightenlines,showpacs,showkeys,notitlepage,nofootinbib]{revtex4-2}

\usepackage[colorlinks=true,citecolor=blue,linkcolor=blue]{hyperref}
\usepackage[normalem]{ulem}
\usepackage{amsmath,amssymb}
\usepackage{mathtools}
\usepackage{verbatim}
\usepackage{titlesec}               
\usepackage{epsfig}
\usepackage{graphicx}               
\usepackage{url}
\usepackage{color}
\usepackage{multirow}
\usepackage{placeins}
\usepackage[dvipsnames]{xcolor}
\usepackage{epstopdf}
\usepackage{siunitx}
\usepackage{fontawesome}
\usepackage{enumitem}
\usepackage[capitalize]{cleveref}
\usepackage{lipsum}
\usepackage{gensymb}
\usepackage{aas_macros}             
\usepackage{booktabs}               
\usepackage{xspace}                 
\usepackage{pifont}
\usepackage{marvosym }
\allowdisplaybreaks

\newcommand{\fstar}{\ding{72}} 
\newcommand{\ostar}{\ding{73}} 

\makeatletter

\renewcommand{\p@subsection}{}
\renewcommand{\p@subsubsection}{}
\makeatother

\titleformat*{\section}{\centering\bfseries\uppercase}
\titlelabel{\thetitle\quad}
\titleformat*{\paragraph}{\bfseries}
\titlespacing*{\paragraph}{0pt}{3.25ex plus 1ex minus .2ex}{1em}

\makeatletter
\def\l@subsubsection#1#2{}
\makeatother

\newcommand{\BR}{\text{BR}}

\newcommand{\iso}[2]{{\ensuremath{{}^{#2}}\ensuremath{\rm #1}}}

\begin{document}

\title{Towards Resolving the Gallium Anomaly}

\author{Vedran Brdar}
\email{vedran.brdar@cern.ch}
\affiliation{Theoretical Physics Department, CERN,
             1 Esplanade des Particules, 1211 Geneva 23, Switzerland}

\author{Julia Gehrlein}
\email{julia.gehrlein@cern.ch}
\affiliation{Theoretical Physics Department, CERN,
             1 Esplanade des Particules, 1211 Geneva 23, Switzerland}

\author{Joachim Kopp}
\email{jkopp@cern.ch}
\affiliation{Theoretical Physics Department, CERN,
             1 Esplanade des Particules, 1211 Geneva 23, Switzerland}
\affiliation{PRISMA Cluster of Excellence \& Mainz Institute for
             Theoretical Physics, \\
             Johannes Gutenberg University, Staudingerweg 7, 55099
             Mainz, Germany}

\date{\today}
\preprint{CERN-TH-2023-036}

\begin{abstract}
\noindent
A series of experiments studying neutrinos from intense radioactive sources have reported a deficit in the measured event rate which, in combination, has reached a statistical significance of $\sim 5\sigma$. In this paper, we explore avenues for explaining this anomaly, both within the Standard Model and beyond. First, we discuss possible biases in the predicted cross section for the detection reaction $\nu_e + \iso{Ga}{71} \to e^- + \iso{Ge}{71}$, which could arise from mismeasurement of the inverse process, \iso{Ge}{71} decay, or from the presence of as yet unknown low-lying excited states of \iso{Ga}{71}. The latter would imply that not all \iso{Ge}{71} decays go to the ground state of \iso{Ga}{71}, so the extraction of the ground state-to-ground state matrix element relevant for neutrino capture on gallium would be incorrect. Second, we scrutinize the measurement of the source intensity in gallium experiments, and we point out that a $\sim 2\%$ error in the branching ratios for \iso{Cr}{51} decay would be enough to explain the anomaly. Third, we investigate the calibration of the radiochemical germanium extraction efficiency as a possible origin of anomaly. Finally, we outline several new explanations beyond the Standard Model, including scenarios with sterile neutrinos coupled to fuzzy dark matter or to dark energy, as well as a model with decaying sterile neutrinos. We critically assess the viability of these scenarios, and others that have been proposed, in a summary table.
\end{abstract}

\maketitle

\tableofcontents

\section{Introduction}
\label{sec:intro}

Neutrino detection using the reaction $\nu_e + \iso{Ga}{71} \to e^- + \iso{Ge}{71}$ has an eventful history: the GALLEX experiment was the first to achieve a measurement of the solar $pp$ neutrino flux in 1992 \cite{GALLEX:1992gcp, Kirsten:2019mli} using this method, followed shortly thereafter by the SAGE experiment \cite{SAGE:1994ctc}. SAGE -- the ``Soviet--American Gallium Experiment'' -- was also seen as demonstrating the power of scientific collaboration in building bridges between politically antipodal countries.  Later, SAGE was caught in the crossfire of the Chechen war, with skirmishes taking place directly outside its host laboratory near Baksan, Russia.  Currently, a new gallium-based neutrino detector is operating in that same laboratory: carrying the assertive name ``BEST'' (Baksan Experiment for Sterile Transitions), the goal of this experiment is to probe a purported $\sim 20\%$ flux deficit of neutrinos from intense radioactive sources.  Such a deficit had been reported in measurements carried out in GALLEX and GNO, albeit with a combined statistical significance of only $\lesssim 3\sigma$ \cite{Anselmann:1994ar, Hampel:1997fc, Abdurashitov:1996dp, Abdurashitov:1998ne, Abdurashitov:2005tb, Giunti:2006bj, Acero:2007su, Kaether:2010ag, Giunti:2010zu, Kostensalo:2019vmv, Kostensalo:2020hbc}. Intriguingly, BEST has recently confirmed this result, dubbed the ``gallium anomaly'', at the $\gtrsim 4\sigma$ level \cite{Barinov:2021asz, Barinov:2021mjj,Barinov:2022wfh, Giunti:2022btk}. The current status of the discrepancy, commonly expressed through the ratio of the number of measured to predicted events, is $R = N_\text{meas}/N_\text{pred} = 0.803\pm 0.035$.

One explanation that has been put forward for the gallium anomaly is mixing between active neutrinos and hypothetical sterile states \cite{Giunti:2006bj, Acero:2007su}. In fact, for some time, reactor experiments appeared to support this hypothesis by reporting a similar deficit \cite{Mueller:2011nm, Mention:2011rk, Huber:2011wv, Gariazzo:2017fdh, Dentler:2017tkw, Dentler:2018sju, Moulai:2019gpi, Giunti:2020uhv, Berryman:2020agd}. More recently, however, it has emerged that the apparent deficit in the reactor neutrino flux was most likely caused by imperfections in the measured beta spectra from nuclear fission that are used as input to reactor neutrino flux calculations \cite{Giunti:2021kab, Berryman:2021yan, Letourneau:2022kfs, STEREO:2022nzk}.

It is therefore more imperative than ever to search for explanations of the gallium anomaly within the Standard Model (SM). This is the main goal of the present paper.  We will discuss several attack vectors, in particular the measured \iso{Ge}{71} decay rate which serves as input to the calculation of the $\nu_e + \iso{Ga}{71}$ cross-section (\cref{sec:nu-capture}), the calorimetric measurement of the source intensity (\cref{sec:Cr51}), and the calibration of the radiochemical germanium extraction efficiency (\cref{sec:calibration}). While, at face value, none of these potential single points of failure can be responsible for the anomaly, our study quantifies the degree to which supporting measurements would need to be off to resolve it. In the second part of the paper (\cref{sec:bsm}), we explore what it would take to explain the gallium anomaly in scenarios beyond the Standard Model (BSM).  We entertain the possibility that resonant active-to-sterile neutrino conversion, driven by sterile neutrino couplings to a fuzzy dark matter (DM) condensate or to dark energy, efficiently depletes electron neutrinos at the energies relevant to gallium experiments using a \iso{Cr}{51} source. In addition, we elaborate on scenarios where sterile neutrinos with eV-scale mass are produced but quickly decay to active neutrinos. We conclude in \cref{sec:conclusions} with a comprehensive summary table that also encompasses other BSM explanations from the literature.

\section{Detection: The Cross Section for Neutrino Capture on Gallium-71}
\label{sec:nu-capture}

Already after the first anomalous measurements of the neutrino capture rate in radioactive source experiments, the cross-section, $\sigma(\nu_e + \iso{Ga}{71})$ for the detection process
\begin{align}
    \nu_e + \iso{Ga}{71} \to e^- + \iso{Ge}{71} \,,
\end{align}
has been called into question \cite{Abdurashitov:2005tb}. $\sigma(\nu_e + \iso{Ga}{71})$ has been thoroughly studied by Bahcall \cite{Bahcall:1997eg} and Haxton \cite{Haxton:1998uc} (see also refs.~\cite{Hata:1995cw, Frekers:2015wga, Kostensalo:2019vmv}), and more recently by Barinov et al.~\cite{Barinov:2017ymq} as well as Semenov \cite{Semenov:2020xea}. There are two contributions: transitions to the ground state of \iso{Ge}{71} (for which the nuclear matrix element is the same as for the well-studied inverse process, electron capture decay of \iso{Ge}{71}) and transitions to excited states of \iso{Ge}{71}, which can only be calculated theoretically, with sizeable uncertainties. Crucially, the anomaly persists even when the latter contribution is set to zero \cite{Giunti:2022btk}.

The most important ingredient for the prediction of $\sigma(\nu_e + \iso{Ga}{71})$ is therefore the measured \iso{Ge}{71} half-life. In the following, we discuss the robustness of this measurement.

\subsection{The Measured Germanium-71 Decay Rate}
\label{sec:Ge71-decay}

The most precise and comprehensive measurement of the \iso{Ge}{71} half-life dates back to Ref.~\cite{Hampel:1985zz}, published in 1985. In that reference, six different measurements were carried out, using two different experimental techniques, and all yielding consistent results. The adopted value for the \iso{Ge}{71} half-life is
\begin{align}
    T_{1/2}(\iso{Ge}{71}) = \SI{11.43 \pm 0.03}{days} \,.
    \label{eq:Hampel}
\end{align}
To fully explain the gallium anomaly, this value would need to be larger by at least \SI{2}{days} ($67 \sigma$), and a reduction of its significance to below $3\sigma$ would still require an increase of $T_{1/2}(\iso{Ge}{71})$ by about one day ($33\sigma$) \cite{Giunti:2022btk}.

\iso{Ge}{71} decays via electron capture from the $K$, $L$, or $M$ shell, so the only observable signal are the X-rays emitted when the electron shell of the nucleus relaxes after the decay, plus possible additional photons from internal bremsstrahlung. As Ref.~\cite{Hampel:1985zz} is very concise, it is difficult to assess in detail the robustness of the measurements reported there.  We outline here a few potential pitfalls that may be worth checking, even though the authors of Ref.~\cite{Hampel:1985zz} were likely aware of them.
\begin{enumerate}
    \item The measurement from Ref.~\cite{Hampel:1985zz}, which is the most precise one to date, disagrees with other, previous measurements. This has been previously emphasized in Ref.~\cite{Giunti:2022xat}. Two publications dating back to the 1950s have reported half-lives of \num{12.5 \pm 0.1} days \cite{Bisi:1955} and \SI{10.5 \pm 0.4}{days} \cite{Rudstam:1956}, respectively,\footnote{We have been unable to access Ref.~\cite{Rudstam:1956}, neither online nor via the University of Uppsala library, therefore we rely here on the results from that reference quoted in the literature.} and a third measurement from 1971 has resulted in a value of \SI{11.15 \pm 0.15}{days} \cite{Genz:1971kv}. These statistically significant discrepancies have never been explained, though the two more recent measurements (Refs.~\cite{Hampel:1985zz, Genz:1971kv}) are in agreement with each other at the $1.9\sigma$ level. It is noteworthy that if we took the half-life from Ref.~\cite{Bisi:1955} at face value, the significance of the gallium anomaly would be reduced to $\lesssim 3\sigma$ \cite{Giunti:2022btk}.

    \item The fact that precision measurements of electron capture decays in general and of \iso{Ge}{71} decay in particular are challenging is highlighted by the reported evidence for a \SI{17}{keV} neutrino in this decay \cite{Zlimen:1991pm}. The infamous and ill-fated \SI{17}{keV} neutrino had first been ``discovered'' in 1985 in tritium beta decay studies \cite{Simpson:1985xc, Morrison:1993}, but the fact that it was falsely ``confirmed'' using the internal bremsstrahlung spectrum in \iso{Ge}{71} decays \cite{Zlimen:1991pm} is certainly a testament to the difficulty of such measurements. (One may argue, though, that a measurement of a decay rate is less error-prone than an investigation of internal bremsstrahlung spectra.)
    
    \item One could speculate whether cosmic rays, or cosmic ray-induced radioactive decays, could mimic the \iso{Ge}{71} decay signal, leading to a spuriously large decay rate measurement. The authors of Ref.~\cite{Hampel:1985zz} comment on possible radioactive impurities, but judge them to be irrelevant. (They do not explicitly comment on radioisotopes produced during the measurement campaign.) Spurious events due to cosmic ray activity would alter the shape of the decay curve. In particular, the extracted instantaneous decay rate would not asymptote to zero in this case. Ref.~\cite{Hampel:1985zz} does not provide enough information to judge if such deviation appears in the data, but it is clear that an effect as large as required to explain the gallium anomaly would not have gone unnoticed.
\end{enumerate}

\subsection{Germanium-71 Decay to New Excited States of Gallium-71?}
\label{sec:Ga71-excitation}

It is generally assumed that electron capture in \iso{Ge}{71} goes to the ground state of the daughter nucleus, \iso{Ga}{71}. In fact, the lowest-lying known excited state of \iso{Ga}{71} has an energy of \SI{389.94 \pm 0.03}{keV}, which is above the $Q$-value of \iso{Ge}{71} decay, \SI{232.49 \pm 0.22}{keV} \cite{Abusaleem:2011exf}.

Here, we speculate on the possibility that there is an additional, yet undiscovered, low-lying excited state of \iso{Ga}{71}. If a $\sim 20$\% fraction of \iso{Ge}{71} went into this state, the nuclear matrix element for ground state-to-ground state transitions (which enters the calculation of $\sigma(\nu_e + \iso{Ga}{71})$) would have been overestimated by the same amount. Correcting for such a bias could resolve the gallium anomaly. Of course, it is unclear how the existence of such an excited state could have been missed, when the state at $\sim \SI{390}{keV}$ has been observed in numerous nuclear reactions, including \iso{Ge}{71} decay \cite{Abusaleem:2011exf}.

\begin{figure}[t]
	\centering
	\includegraphics[scale=0.65]{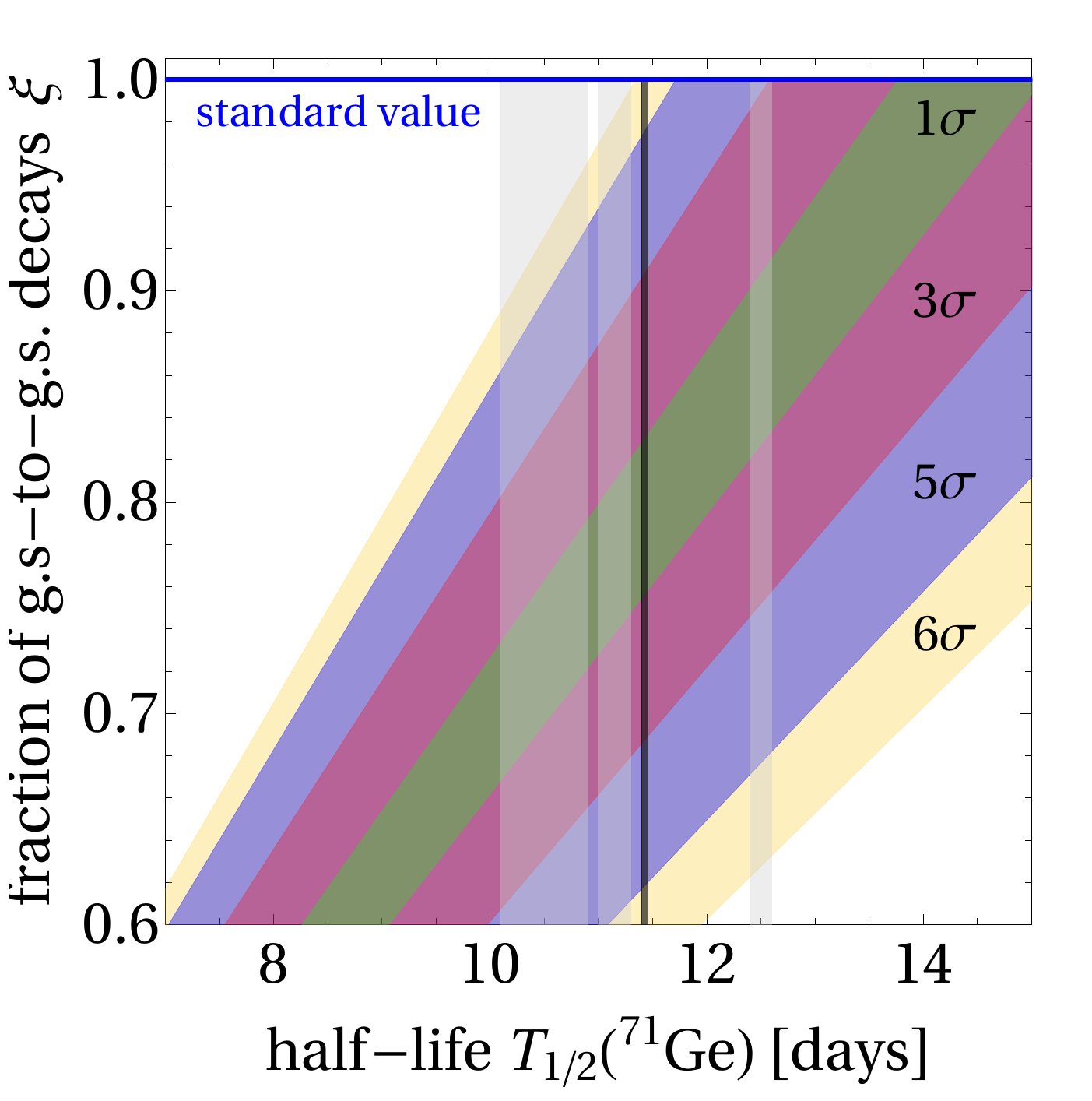} 
	\caption{Statistical significance of the gallium anomaly as a function of the \iso{Ge}{71} half-life, $T_{1/2}(\iso{Ge}{71})$, and the fraction of \iso{Ge}{71} decays ending up in the ground state of \iso{Ga}{71}, $\xi$. The $1\sigma$, $3\sigma$, $5\sigma$, and $6\sigma$ regions are shown as colored bands. The measurement of $T_{1/2}(\iso{Ge}{71})$ from Ref.~\cite{Hampel:1985zz} is included as a black vertical line, and the $1\sigma$ confidence intervals from previous measurements of this quantity are shown in gray (see text for a discussion of the discrepancies between these measurements). For parameter points within the green band, the anomaly is fully resolved; we see that this would require either $\xi$ to be $\sim 20$\% smaller than previously thought, or the measured half-life to be off by a similar amount (2--3\,days).}
	\label{fig:T12-xi}
\end{figure}

In \cref{fig:T12-xi} we show the statistical significance of the gallium anomaly as a function of $T_{1/2}(\iso{Ge}{71})$ (the \iso{Ge}{71} half-life) and $\xi$ (the fraction of \iso{Ge}{71} decays that go to the ground state of \iso{Ga}{71}). Our estimates are based on the cross-section calculations by Bahcall \cite{Bahcall:1997eg}, and on the statistical procedure proposed by Giunti et al.~in Ref.~\cite{Giunti:2022xat} (with a small modification described below).  More precisely, we define the test statistic
\begin{align}
    \chi^2(\bar{R}) \equiv \min_\eta \bigg[ \sum_\text{exp} \bigg(
                           \frac{R_\text{exp} - \eta \bar{R}}{\Delta R_\text{exp}} \bigg)^2
                         + \chi_\eta^2 \bigg] \,,
    \label{eq:chi2}
\end{align}
where $\bar{R}$ is the fitted ratio between the observed and predicted event rates, $R_\text{exp}$ is the measured value of this ratio in a particular experiment denoted by the subscript ``exp'', $\Delta R_\text{exp}$ is the corresponding uncertainty, and $\text{exp} \in \{ \text{GALLEX-1},\allowbreak \text{GALLEX-2},\allowbreak \text{SAGE-Ar},\allowbreak \text{SAGE-Cr},\allowbreak \text{BEST-inner},\allowbreak \text{BEST-outer}\}$.  The nuisance parameter $\eta$ accounts for the cross-section uncertainty which is correlated between experiments. The philosophy here is that $\bar{R}$ is the ratio of measured and predicted event rates in a particular scenario (possibly including new effects like the ones described here, or BSM physics), normalized to the SM prediction assuming the central value from the Bahcall model for the neutrino--gallium scattering cross-section. $\eta$ parameterizes the wiggle room in the predicted event rate afforded by the cross-section uncertainty. Note that the values for $\Delta R_\text{exp}$ quoted by the experimental collaborations (except SAGE-Cr) include the systematic uncertainty in the neutrino--gallium cross-section (based on Bahcall's cross-section calculation). As we include this uncertainty separately through $\eta$ we need to remove it from $\Delta R_\text{exp}$ to avoid double-counting. We do so by replacing $\Delta R_\text{exp}$ with $[\Delta R_\text{exp}^2 - (R_\text{exp} \times \Delta\sigma_\text{Bahcall} / \sigma_\text{Bahcall})^2 ]^{1/2}$. The nuisance parameter $\eta$ is constrained by the pull term%
\footnote{While this is a consistent approach in searches for BSM physics, one could argue that it is not quite appropriate when we consider modifications to the event rate due to changes in $T_{1/2}(\iso{Ge}{71})$ or $\xi$ because by postulating such changes we effectively claim that the estimate for the amount of wiggle room, $\Delta\sigma_\text{Bahcall}$, which enters \cref{eq:chi2-eta} is incorrect.}
\begin{align}
    \chi_\eta^2 \equiv \begin{cases}
                           \Big( \frac{(1-\eta) \sigma_\text{Bahcall}}{\Delta \sigma_\text{Bahcall}} \Big)^2
                               &  \text{for $\eta \geq \eta_\text{crit}$} \,, \\
                           \Big( \frac{\sigma_\text{Bahcall} - \sigma_\text{gs}}
                                      {\Delta \sigma_\text{Bahcall}} \Big)^2
                         + \Big( \frac{(1-\eta) \, \sigma_\text{Bahcall}
                                           - (\sigma_\text{Bahcall} - \sigma_\text{gs})}
                                       {\Delta \sigma_\text{gs}} \Big)^2
                               &  \text{for $\eta < \eta_\text{crit}$} \,. \\
                       \end{cases}
    \label{eq:chi2-eta}
\end{align}
The first line of this expression is a standard pull term that constrains the nuisance parameter to remain within the stated uncertainty, $\Delta \sigma_\text{Bahcall}$, of the cross section calculated by Bahcall, $\sigma_\text{Bahcall}$. This uncertainty is dominated by the uncertainty in the excited-state cross-section $\sigma(\nu + \iso{Ga}{71} \to e^- + \iso{Ge}{71}^*)$.  However, since this contribution to the total cross-section cannot be less than zero, it is inappropriate to use its uncertainty at small $\eta$. In that regime, the first term in the second line of \cref{eq:chi2-eta} accounts for the bias in the excited-state cross-section needed to set it to zero, and the second term constrains any additional bias to be within the much smaller uncertainty $\Delta\sigma_\text{gs}$ of the ground-state cross-section, $\sigma_\text{gs}$. The critical value, $\eta_\text{crit}$, is determined by equating the expressions in the first and second line of \cref{eq:chi2-eta}. It is simply $\eta_\text{crit} = \sigma_\text{gs} / \sigma_\text{Bahcall}$. The numerical values of the parameters appearing here are $\sigma_\text{gs} = \SI{5.539e-45}{cm^2}$, $\Delta \sigma_\text{gs} = \SI{0.019e-45}{cm^2}$, $\sigma_\text{Bahcall} = \SI{5.81e-45}{cm^2}$, $\Delta \sigma_\text{Bahcall} = \SI{0.16e-45}{cm^2}$ \cite{Giunti:2022xat}.

In order to account for a possible bias in the measurement of $T_{1/2}(\iso{Ge}{71})$ and for the possible existence of excited states in \iso{Ga}{71}, we rescale both $R_\text{exp}$ and $\Delta R_\text{exp}$ in \cref{eq:chi2} by a factor
\begin{align}
    (1 + 0.051) \times \bigg(0.051 + \xi \, \frac{T_{1/2}^\text{HR}}{T_{1/2}(\iso{Ge}{71})} \bigg)^{-1}\,,
    \label{eq:rescale_1}
\end{align}
where $0.051$ represents the contribution from the transition into the excited states of \iso{Ge}{71} (this number is also obtained in \cite{Giunti:2022xat}) and $T_{1/2}^\text{HR}$ is the standard value of the germanium lifetime from the measurements by Hampel and Remsberg~\cite{Hampel:1985zz}, quoted in \cref{eq:Hampel}.

\Cref{fig:T12-xi} confirms that, with the most recent measurements of $T_{1/2}(\iso{Ge}{71})$ \cite{Hampel:1985zz} and with the standard assumption $\xi=1$, the significance of the gallium anomaly is $\sim 5\sigma$.  The plot also illustrates the conclusions from \cref{sec:Ge71-decay}, namely that the statistical significance would be reduced to below $1\sigma$ if the true half-life were larger, $T_{1/2}(\iso{Ge}{71}) \sim \SI{14}{days}$. Finally, we clearly observe how a reduction of $\xi$ has the same effect.  If only 80\% of the \iso{Ge}{71} decays contributing to the event sample in Ref.~\cite{Hampel:1985zz} were ground state-to-ground state decays, the gallium anomaly could be explained. 

Given that the main purpose of the gallium experiments GALLEX and SAGE was the measurement of the solar $pp$ neutrino flux at the 10\% level, a reduction of the neutrino capture cross section on gallium would impact this measurement as well~\cite{SAGE:2002fps, SAGE:2009eeu, Gonzalez-Garcia:2009dpj, Bergstrom:2016cbh}. In particular, a decrease of the cross section by $\sim 20$\%, as suggested by the gallium anomaly, would lead to a commensurate increase in the extracted $pp$ neutrino flux, in potential tension with the tight constraints on this flux from the observed total luminosity of the Sun \cite{Bergstrom:2016cbh}.  However, the tension would only be at the $\sim 2\sigma$ level, so solar neutrino measurements are not in statistically significant conflict with the possibility that the neutrino capture cross section on \iso{Ga}{71} is smaller by $\sim 20$\%. This conclusion does not change when the more recent measurement of the solar $pp$ neutrino flux by Borexino is included \cite{BOREXINO:2018ohr}.

\section{Source: Chromium-51 Branching Ratios}
\label{sec:Cr51}

A second crucial ingredient in the experiments at the origin of the gallium anomaly is the prediction of the neutrino flux emitted by the source. Most experiments to date have employed a \iso{Cr}{51} source, which is produced by neutron irradiation of chromium metal enriched in \iso{Cr}{50} \cite{Danshin:2022wiz}. \iso{Cr}{51} decays via electron capture,
\begin{align}
    \iso{Cr}{51} + e^- \to \iso{V}{51} + \nu_e \,,
\end{align}
with a half-life of \SI{27.704 \pm 0.004} days \cite{Wang:2017jnh}. Only SAGE has also used an \iso{Ar}{37} source (electron capture decay to \iso{Cl}{37}, $T_{1/2} = \SI{35.011 \pm 0.019}{days}$ \cite{Cameron:2012ogv}). However, this measurement plays only a subdominant role in the overall evidence for a neutrino deficit and hence we focus here on \iso{Cr}{51} sources.

The source intensity in these experiments is measured calorimetrically \cite{Gavrin:2021lam}, and since the decay is via electron capture, the main heat sources are X-rays from the de-excitation of the electron shell and gamma rays from the de-excitation of the daughter nucleus, with the latter contribution dominant by far. In fact, $\sim 10$\% of all \iso{Cr}{51} decays populate the first excited state of \iso{V}{51} at \SI{320.0835 \pm 0.0004}{keV} instead of the ground state. And as almost all the heat production comes from the $\sim \SI{320}{keV}$ de-excitation gamma rays, we see that the measurement of the source intensity is based on only $\sim 10\%$ of all decays.  In other words, if the true branching ratio for decays to the excited state, $\BR^{\text{exc}} \equiv \BR(\iso{Cr}{51} \to \iso{V}{51}^*)$, was larger by only $\sim 2$\%, the source intensity would have been overestimated by $\sim 20$\%, enough to explain the gallium anomaly.

\begin{figure}[t]
	\centering
	\includegraphics[scale=0.65]{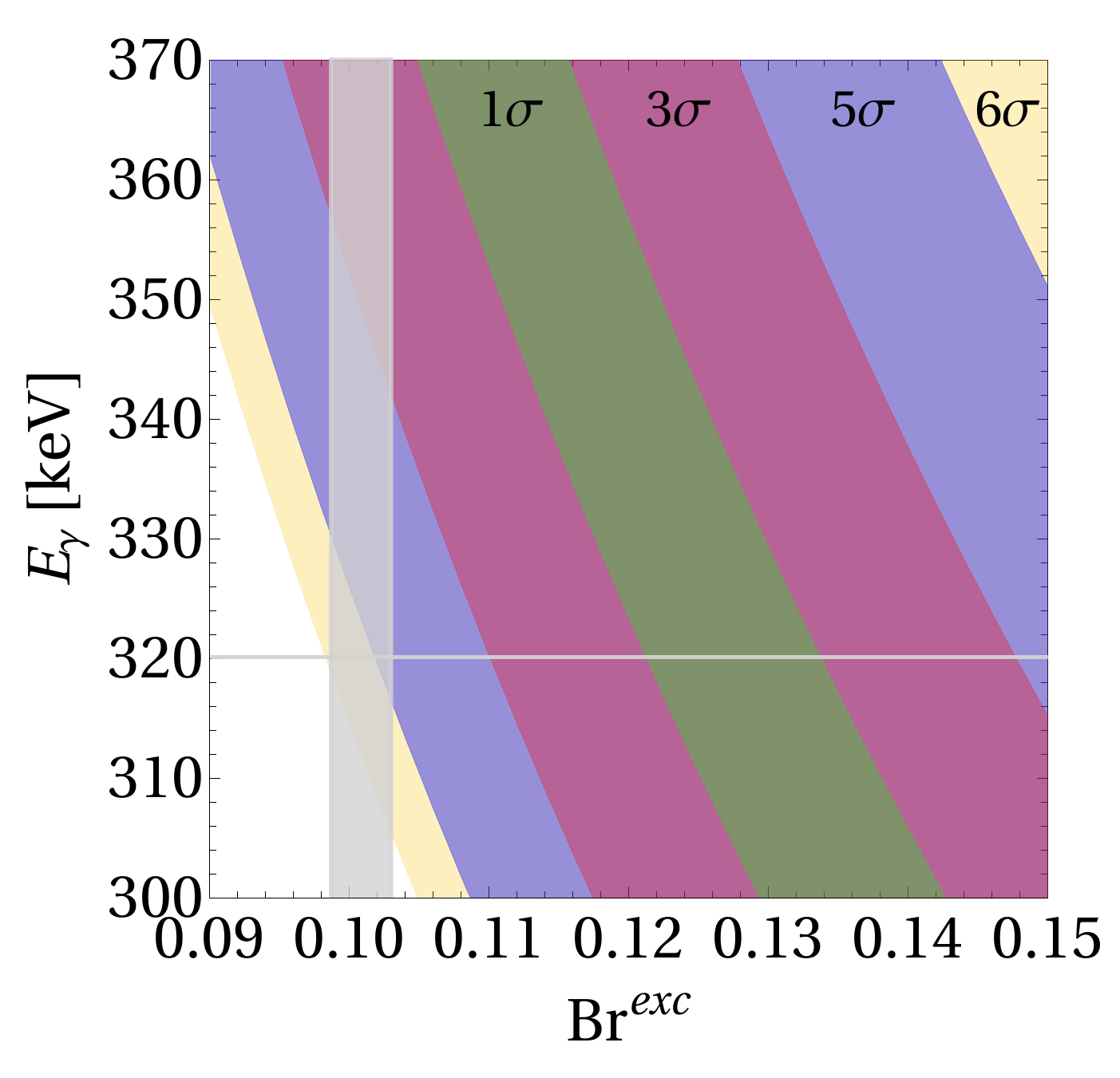} 
	\caption{Statistical significance of the gallium anomaly as a function of $\BR^\text{exc}$, (the branching ratio for \iso{Cr}{51} decay into the lowest-lying excited state of \iso{V}{51}), and the energy of the de-excitation gamma rays, $E_\gamma$. The $1\sigma$, $3\sigma$, $5\sigma$, and $6\sigma$ regions are shown as colored bands. The measured values for both parameters (with $1\sigma$ uncertainties) are shown in gray.}
	\label{fig:BRexc-Egamma}
\end{figure}

This is illustrated in \cref{fig:BRexc-Egamma}, which shows the significance of the anomaly as a function of $\BR^\text{exc}$ on the horizontal axis, and the energy of de-excitation gamma, $E_\gamma$, on the vertical axis. This plot is based on an analysis similar to the one underlying \cref{fig:T12-xi} above (see again \cref{eq:chi2}), but with $R_\text{exp}$ and $\Delta R_\text{exp}$ rescaled by
\begin{align}
    \frac{\BR^\text{exc} E_\gamma + \SI{5.03}{keV}}{\SI{36.75}{keV}}\,.
    \label{eq:source}
\end{align}
In the denominator of this expression, \SI{36.75}{keV} is the average total visible energy output per \iso{Cr}{51} decay, including X-rays, internal bremsstrahlung, and gamma rays, and accounting for the relative probabilities for $K$, $L$, and $M$ capture, as well as the branching ratio, $\BR^\text{exc}=0.0991$, into the excited state of $\iso{V}{51}^*$ at $E_\gamma = \SI{320.0835}{keV}$ \cite{Veretenkin_2017}. In the numerator, the first term describes the contribution to the visible energy from $\iso{V}{51}^*$ de-excitation under the assumption of modified $\BR^\text{exc}$ and $E_\gamma$, while the second term (\SI{5.03}{keV}) describes the energy release from X-rays and internal bremsstrahlung, which is present in all \iso{Cr}{51} decays, including those to the ground state of \iso{V}{51}.

At the established values $E_\gamma \simeq \SI{320}{keV}$ and $\BR^\text{exc} \sim 10\%$, indicated by gray bands in \cref{fig:BRexc-Egamma}, the anomaly is at $\sim 5\sigma$ significance. The figure confirms that increasing $\BR^\text{exc}$ to $\sim 12$--$13\%$ would lower the significance to below $1\sigma$. We note that different measurements of $\BR^\text{exc}$ differ by $0.5$\% \cite{Barinov:2022wfh} (as shown by the width of the gray vertical band in \cref{fig:BRexc-Egamma}), but are consistent with each other at the $3\sigma$ level, see Ref.~\cite{Wang:2017jnh} and references therein. Pursuing the hypothesis of a branching ratio mismeasurement further, one would therefore need to explain why more than ten different measurements should show the same systematic bias.

Another avenue for attaining larger energy per decay is increasing $E_\gamma$ while keeping $\BR^\text{exc} \simeq 10$\%; if the energy of the excited state was \SI{360}{keV} instead of \SI{320}{keV}, the significance of the anomaly would drop from $\sim 5\sigma$ to $\sim 3\sigma$. But since measurements of nuclear excitation energies are rather robust, the only reasonable way to achieve such a large shift would be to postulate the existence of another, yet undiscovered, excited state in \iso{V}{51} with an energy larger than \SI{320}{keV}, but below the $Q$-value of $\iso{Cr}{51}$ decay, \SI{752.39}{keV}. In fact, there is weak evidence for a state at \SI{470}{keV} \cite{Wang:2017jnh, Maheshwari:1971}, with hints for it observed also in \iso{Cr}{51} decay \cite{Mathe:1963}. However, these observations from the 1960s and 70s have not found further support in more recent measurements. Moreover, Ref.~\cite{Mathe:1963} suggests that the relative intensity of the \SI{470}{keV} line is $< \num{e-5}$, which would be too faint to explain a 2--3\% bias in the calorimetric source intensity measurement in the gallium experiments. Nevertheless, the potential relevance of an extra excited state for our understanding of the gallium anomaly calls for a renewed effort to establish or conclusively refute its existence.

An alternative avenue to increase $E_\gamma$ is to envision a new heat source that is presently unaccounted for. But while it may be possible to conceive an extension of the SM featuring such a heat source in \iso{Cr}{51} decay, it would most likely also impact other radioactive decays.  Given the amount of energy that would need to go into this new channel, it is hard to imagine that it would have been missed so far. For one, a new heat source in nuclear beta decays would impact the energy output of nuclear reactors. Typically, around 6.5\% of the total heat in a reactor originates from beta decay of fission products. The additional heat required to explain the gallium anomaly would constitute an $\approx 0.5\%$ correction to this. Given the long history of reactor monitoring, and the industry's excellent understanding of reactor cores, such an additional heat source would most likely have been noticed. Furthermore, possible BSM realizations in which photons carry the additional heat (such as a sterile neutrino which decays to a photon and an active neutrino via the transition magnetic moment portal \cite{Brdar:2020quo}) suffer from strong constraints.

\section{Calibration of the Radiochemical Germanium Extraction Efficiency}
\label{sec:calibration}

If any experimental technique in the history of physics deserves to be described as a search for a needle in a haystack, it is certainly radiochemical neutrino detection. Extracting $\mathcal{O}(100)$ \iso{Ge}{71} nuclei from more than \SI{47}{tons} of liquid gallium, as done in BEST, seems to be a formidable task. Nevertheless, the GALLEX, SAGE, and BEST collaborations have rather convincingly demonstrated their ability to pull off this feat. In particular, SAGE \cite{Cleveland:2015tlq} and BEST \cite{Barinov:2022wfh} have deliberately added small ($\lesssim \mathcal{O}(\si{\micro mol})$) amounts of stable germanium with well-defined isotope ratios to the detection volume. The amount of stable Ge, while tiny, still exceeds the number of \iso{Ge}{71} nuclei produced in neutrino interactions by many orders of magnitude.  After each run of the experiment, the extracted germanium is studied via mass spectrometry to verify that the amount and the isotope ratios of the extracted stable germanium match the known properties of the germanium that was added to the detector before the run. As the chemical behavior of different Ge isotopes should be identical, the extraction efficiency determined this way (which is close to 100\%) can be assumed to hold also for unstable \iso{Ge}{71}.

A second indication that germanium extraction seems to be well understood is the fact that measurements of the solar neutrino flux in GALLEX and SAGE agree with other solar neutrino measurements. However, as we discussed at the end of \cref{sec:nu-capture}, such measurements still suffer from $\mathcal{O}(10\%)$ uncertainties; this means that an overestimation of the germanium extraction efficiency by this amount would not be excluded by solar data.

\begin{figure}[t]
	\centering
	\includegraphics[scale=0.35]{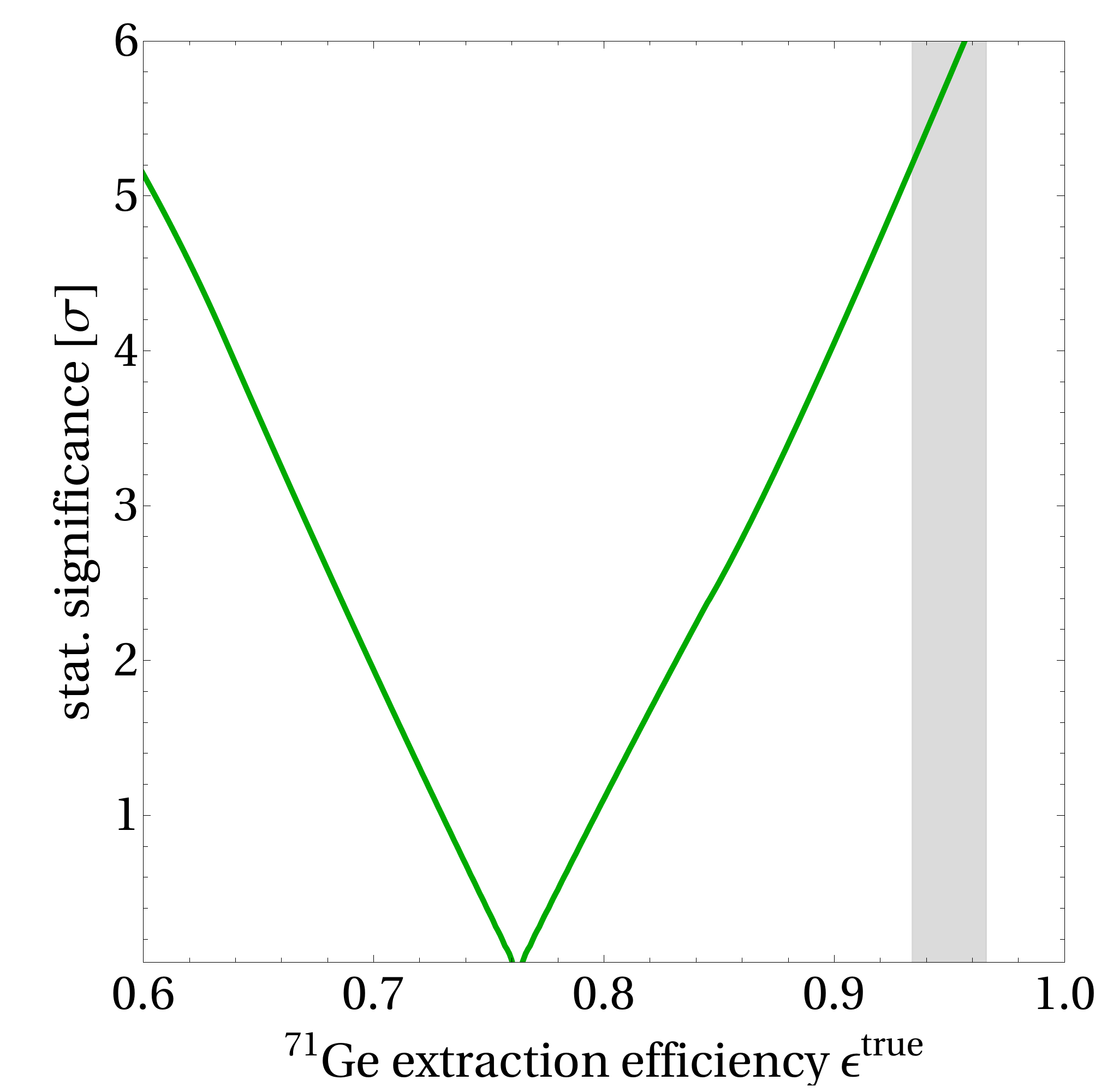} 
	\caption{Statistical significance of the gallium anomaly as a function of the \iso{Ge}{71} extraction efficiency, $\epsilon^\text{true}$. The gray band indicates the nominal extraction efficiency and its standard deviation, $\epsilon^\text{cal} = 0.95 \pm 0.016$, based on calibration measurements in BEST \cite{Barinov:2022wfh}.}
	\label{fig:epsilon}
\end{figure}

In fact, if the true extraction efficiency $\epsilon^\text{true}$ was different from the estimate from calibration measurements, $\epsilon^\text{cal}\approx 0.95$, the ratio of the observed to predicted event rate in gallium experiments would change by a factor
\begin{align}
    \frac{\epsilon^\text{cal}}{\epsilon^\text{true}}\,.
    \label{eq:ratio_cal}
\end{align}
To account for such a bias, we rescale $R_\text{exp}$ and $\Delta R_\text{exp}$ in \cref{eq:chi2} by the factor in \cref{eq:ratio_cal}. The result is illustrated in \cref{fig:epsilon}, which shows the statistical significance of the anomaly as a function of $\epsilon^\text{true}$. It is clear that resolving the anomaly would require the calibration to be off by around 20\%, namely $\epsilon^\text{true}\sim 0.75$. Barring an analysis error, this would imply the existence of an unidentified route for (stable) germanium to enter the detector. The amount of extra germanium needed may seem small ($< \si{\micro mol}$), but still corresponds to $\sim 10^{17}$ atoms. This is more than can be realistically produced by cosmic rays for instance. (Even though a route for cosmic ray-induced production of a small number of stable germanium atoms may exist via neutron capture on \iso{Ga}{71}, followed by $\beta^-$ decay to stable \iso{Ge}{72}.)

Note however that in one extraction of the SAGE experiment, an anomalously large amount of stable Ge (about 10 times larger than expected) was found \cite{Cleveland:2015tlq}. At the time, this result was attributed to an unidentified experimental accident and the corresponding data point was discarded. But no source for the apparent contamination has been identified. Nevertheless, this occurrence illustrates that there appear to be pathways for sizable amounts of extra germanium to enter a liquid gallium detector. We cannot rule out the possibility that such contamination (albeit at a smaller level) has occurred in many runs of the SAGE and BEST experiments, and that the germanium extraction efficiency has therefore been overestimated.

\section{Ideas Beyond the Standard Model}
\label{sec:bsm}

In the previous sections we have discussed several possible explanations for the ``gallium anomaly'' within the SM. Nevertheless, the observed deficit and its high statistical significance could also provide an intriguing hint for new physics. The ``usual'' BSM interpretation of the gallium anomaly features eV-scale sterile neutrinos which mix with electron neutrinos \cite{Giunti:2006bj, Acero:2007su}. However, given that the large mixing angle required in this scenario is heavily constrained by solar and reactor data  \cite{Giunti:2022btk, Berryman:2021yan} it is tempting to explore other possibilities. In this section we will investigate several such scenarios which have not been discussed in the literature before. We will begin in \cref{sec:bsm-res} with a model in which oscillations between active and sterile states are stimulated by a Mikheyev-Smirnov-Wolfenstein (MSW) resonance \cite{Wolfenstein:1977ue, Mikheyev:1985zog} or a parametric resonance \cite{Petcov:1998su, Akhmedov:1998ui}. This will be followed by a decaying sterile neutrino scenario in \cref{sec:bsm-nu-s-decay}.  Other new physics efforts that have been proposed over the last decade will be brought up in \cref{sec:conclusions} where we will present a summary table.

\subsection{Sterile Neutrinos Coupled to Fuzzy Dark Matter or Dark Energy}
\label{sec:bsm-res}

Solar and reactor constraints on the mixing between active and sterile neutrinos do not allow for the $\mathcal{O}(20\%)$ $\nu_e$ disappearance that would be required to explain the gallium anomaly. Avoiding these constraints is challenging because neutrinos produced in the Sun and in nuclear reactors have energies similar to those emitted from \iso{Cr}{51} decay. In the following, we will circumvent this problem by invoking a sharp resonance -- either an MSW resonance or a parametric resonance -- whose position coincides with two of the four neutrino lines in \iso{Cr}{51} decay. The energies and relative intensities of these lines are \SI{747}{keV} (81.63\%), \SI{427}{keV} (8.95\%), \SI{752}{keV} (8.49\%), and \SI{432}{keV} (0.93\%) \cite{Acero:2007su}.  (For the \iso{Ar}{37} source used in some SAGE runs, the neutrino energies and line intensities are \SI{811}{keV} (90.2\%) and \SI{813}{keV} (9.8\%) \cite{Endt:1990zz}). These energies are above the cutoff of the solar \textit{pp} neutrino flux at $\sim \SI{420}{keV}$, but below the energy of solar \iso{Be}{7} neutrinos at \SI{861.8}{keV}; other components of the solar neutrino flux are subdominant in this energy range. Further, neutrinos with sub-MeV energies cannot be recorded via inverse beta decay, which is the typical detection channel for reactor neutrinos. Therefore, if active--sterile mixing is locally enhanced around the neutrino energies emitted by \iso{Cr}{51}, there may be no observable effect at other neutrino experiments.

\subsubsection{MSW resonance from Interaction with Fuzzy Dark Matter}
\label{subsec:MSW}

A straightforward way to realize a sharp resonance is the MSW effect which neutrinos experience as they travel through background matter \cite{Wolfenstein:1977ue, Mikheyev:1985zog}. Working in the effective two flavor (electron neutrino $\nu_e$ and sterile neutrino $\nu_s$) framework, the mixing angle in matter, $\theta_{e4}^\text{eff}$, is given by
\begin{align}
  \sin 2 \theta_{e4}^\text{eff} =
    \frac{\frac{\Delta m^2}{2 E_\nu} \sin 2\theta_{e4}^\text{vac}}
         {\sqrt{(V - \frac{\Delta m^2}{2 E_\nu} \cos 2\theta_{e4}^\text{vac})^2
           + (\frac{\Delta m^2}{2 E_\nu})^2 \sin^2 2\theta_{e4}^\text{vac}}} \,,
  \label{eq:MSW}
\end{align}
where $\theta_{e4}^\text{vac}$ is the vacuum mixing angle, $\Delta m^2$ is the mass squared difference between the two neutrino mass eigenstates in vacuum, $E_\nu$ is the neutrino energy, and $V$ is the difference between the matter potential felt by $\nu_e$ and $\nu_s$. The mixing angle is resonantly enhanced for
\begin{align}
  V = \frac{\Delta m^2}{2 E_\nu^\text{res}} \cos 2\theta_{e4}^\text{vac} \,,
  \label{eq:res}
\end{align}
as can be easily seen from \cref{eq:MSW}.

Unfortunately, the SM matter potential generated by the weak interaction does not suffice to achieve close-to-maximal active--sterile mixing at energies as low as those probed by gallium experiments.  It is suppressed by the Fermi constant, $G_F$, which induces MSW resonances only at higher energies. We therefore need to introduce a new interaction, which we choose to be an interaction with an ultralight DM particles, in particular an ultra-light vector boson $\phi^\mu$ \cite{Nelson:2011sf, Brdar:2017kbt}.%
\footnote{Couplings to an ultralight scalar do not lead to a new MSW resonance, but rather to an energy-independent correction to neutrino masses and mixing angles \cite{Berlin:2016woy, Krnjaic:2017zlz, Brdar:2017kbt}, see also next subsection.}
The relevant part of the Lagrangian reads
\begin{align}
  \mathcal{L}\supset  g_s \phi^\mu \overline{\nu}_s \gamma_\mu \nu_s \,.
  \label{eq:lag}
\end{align}
Here, $\nu_s$ is the sterile neutrino field and $g_s$ is a dimensionless coupling constant. The interaction in \cref{eq:lag} leads to an effective MSW potential \cite{Brdar:2017kbt}
\begin{align}
  V = -\frac{1}{2E_\nu} \left(2 (p_\nu \cdot \phi) g_s + g_s^2\phi^2\right) \,.
  \label{eq:mswdark}
\end{align}
We treat $\phi^\mu$ as a classical field which coherently oscillates in time, i.e.\ $\phi^\mu = \phi_0 \xi^\mu \cos(m_\phi t)$, where $\xi^\mu$ is a polarization vector. We assume that the DM field is uniformly polarized over macroscopic regions of at least several meters (the size of the gallium experiments), see ref.~\cite{Brdar:2017kbt} for a discussion of this assumption. Without polarization, fuzzy vector DM would affect neutrino oscillations in the same way as scalar DM, that is, it would not induce an MSW resonance, but merely an energy-independent correction to neutrino masses. We discuss such a scenario in the next subsection. It is crucial that the $\cos(m_\phi t)$ term has remained nearly constant over the last $\sim 30$ years when gallium experiments have been operating (for simplicity, we will use $\cos(m_\phi t) = 1$ in the following; for smaller values, the coupling constant $g_s$ would need to be rescaled accordingly); otherwise, detuning of the MSW resonance would have occurred, moving it away from the \iso{Cr}{51} emission line. This imposes the constraint $m_\phi \lesssim \SI{e-24}{eV}$, which appears to be in conflict with, for instance, Lyman-$\alpha$ forests \cite{Kobayashi:2017jcf, Rogers:2020ltq}. To evade this type of constraint while keeping $m_\phi$ low, we assume that fuzzy DM constitutes only $\sim 1$\% of the total DM energy density in the Universe. Lyman-$\alpha$ constraints become insensitive if the fuzzy DM fraction drops below 0.2 for $m_\phi < \SI{e-22}{eV}$ \cite{Kobayashi:2017jcf}, and other large scale structure-related observables sensitive to the fuzziness of DM should weaken in a similar way. Given that the local DM density is $\SI{0.3}{GeV/cm^3}$ \cite{ParticleDataGroup:2022pth}, we hence take $\rho_\phi \simeq \SI{0.003}{GeV/cm^3}$.

The oscillation amplitude of the fuzzy DM field is given by
\begin{align}
  \phi_0 = \frac{\sqrt{2\rho_{\phi}}}{m_\phi} \,.
\end{align}
We choose the sign of the sterile neutrino--DM coupling, $g_s$, such that the new MSW resonance lies in the neutrino sector, while antineutrino mixing is never resonantly enhanced. As a result, neutrinos produced at gallium experiments as well as solar neutrinos could experience enhancements, as illustrated in \cref{eq:MSW}, while antineutrinos would not be significantly affected.

The interaction between ultralight DM and active neutrinos in the context of neutrino oscillations has been extensively studied for instance in Refs.~\cite{Berlin:2016woy, Krnjaic:2017zlz, Brdar:2017kbt, Alonso-Alvarez:2023tii,Huang:2022wmz}. In Ref.~\cite{Brdar:2017kbt}, bounds on ultralight vector DM coupling to active neutrinos were derived using accelerator neutrino data from T2K, solar as well as reactor data. For the case considered here, where DM only couples to sterile neutrinos, the situation is more involved. First, at baselines where standard oscillations matter, we can no longer work in the two-flavor approximation. Instead, let us generalize the above scenario to three sterile neutrinos, which we take to be mass-degenerate for simplicity. We assume each sterile neutrino to mix with exactly one of the mostly active mass eigenstates, and that the three corresponding mixing angles are all identical. Finally, we take the couplings of the three sterile neutrinos to DM to be identical.\footnote{With only one sterile neutrinos, or with multiple sterile neutrinos whose mixing structure is not the same as the one of the SM neutrinos, the effective mixing angles between active neutrinos would be distorted. This may not be a fundamental problem as the changes could be absorbed into a redefinition of these mixing angles as long as the DM field is constant over the time and distance scales over which neutrino oscillation experiments have been carried out.} In this case, constraints from terrestrial experiments are significantly alleviated.  This is because we are interested in an MSW resonance that is strongly peaked in energy. This implies small $\theta_{e4}^\text{vac}$ as the resonance width is given by $\Delta E_\nu^\text{res} = E_\nu^\text{res} \tan 2\theta_{e4}^\text{vac}$ \cite{Wolfenstein:1977ue, Mikheyev:1985zog}. With a small vacuum mixing angle, active-to-sterile neutrino oscillations are suppressed away from the resonance, which renders reactor and accelerator experiments, which probe energies higher than the ones accessible in \iso{Cr}{51} decay, insensitive to the existence of the sterile neutrino.  In the case of accelerator constraints, there is extra suppression because at energies far above the MSW resonance, the denominator of \cref{eq:MSW} is $\approx V$ and the effective mixing angle, $\sin2\theta_{e4}^{\text{eff}}$, is further reduced by a factor $\simeq \Delta m_{41}^{2,\text{vac}}/(2 E_\nu V)$ compared to $\sin2\theta_{e4}^{\text{vac}}$.

Similar arguments can be made for solar neutrinos: the high-energy part of the spectrum remains largely unperturbed thanks to the small $\theta_{e4}^{\text{vac}}$, supported by moderate MSW suppression. Low-energy $pp$ neutrinos, whose energy is below the MSW resonance, still benefit from the small mixing angle in vacuum $\theta_{e4}^{\text{vac}}$.  The most critical component are the \iso{Be}{7} neutrinos, from which a limit of $\theta_{e4}^\text{vac} \lesssim 4.5^\circ$ can be derived, based on the requirement that the resonance should be sufficiently narrow not to affect the \iso{Be}{7} lines: $\Delta E_\nu^\text{res} < E_{\iso{Be}{7}} - E_{\iso{Cr}{51}} \approx \SI{120}{keV}$.  These arguments are based on the assumption that the DM density in the Sun is similar to the one in the rest of the solar system, so that no adiabatic flavor transitions occur as neutrinos travel out of the Sun. This is reasonable based on the weakness of the ``gravitational focusing'' effect that the Sun's gravitational field has on the local DM density (see Ref.~\cite{Lee:2013wza} for the case of ordinary WIMP DM, and Ref.~\cite{Kim:2021yyo} for fuzzy DM).%
\footnote{This argument would be invalidated if DM had substantial self-interactions, or interactions with ordinary matter, as in this case the Sun could capture DM and accumulate it at its center.}
If there was an overdensity of DM inside the Sun, adiabatic conversion would still affect only $pp$ neutrinos. This is because at higher density, the resonance moves to \emph{lower} energies. $pp$ neutrinos could then cross it, while all other components of the solar neutrino flux have energies that are above the resonance both inside and outside the Sun.

For these reasons, we can use the solar neutrino constraint on $\theta_{e4}^\text{vac}$ from the literature at face value, assuming no additional resonances in the Sun.  The correction to the (high-energy) solar electron neutrino survival probability due to the existence of a fourth neutrino flavor is $\propto (\theta_{e4}^\text{eff})^4$ \cite{Brdar:2017kbt}, which for $\theta_{e4}^\text{vac} \lesssim 5^\circ$ is at the sub-per mille level, far below experimental sensitivities. For example, the \iso{Be}{7} neutrino flux, with an energy closest to the neutrinos from \iso{Cr}{51}, has been measured with an impressive, but in this case insufficient, uncertainty of 3.5\% \cite{BOREXINO:2018ohr}. The survival probability of lower-energy solar $pp$ neutrinos is only modified by terms of order $(\theta_{e4}^\text{eff})^2$, but this is still small enough to be compatible with current experiments. 

\begin{figure}[t]
	\centering
	\includegraphics[scale=0.7]{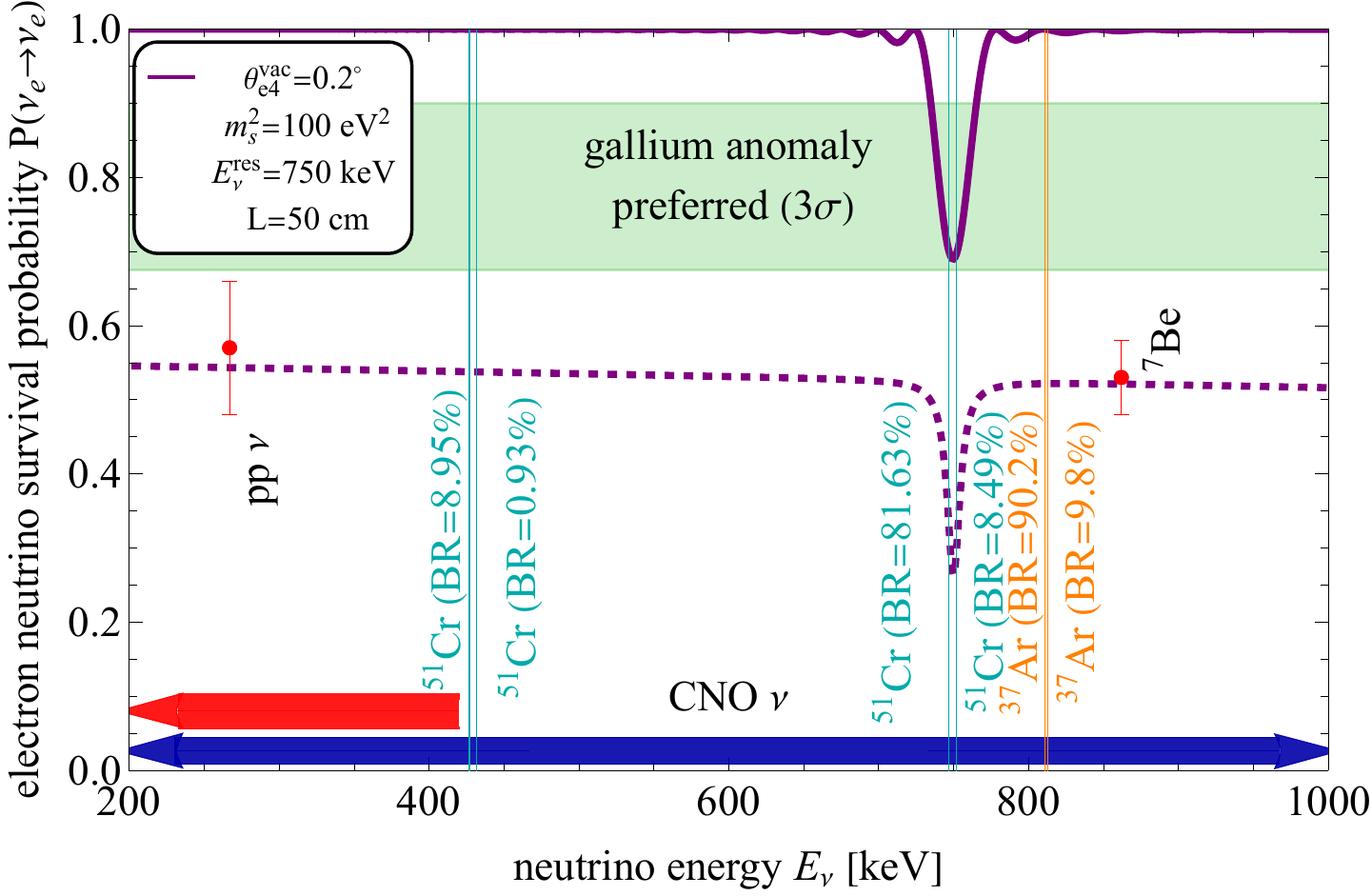} 
	\caption{The electron neutrino survival probability as a function of energy for the scenario from \cref{subsec:MSW}, where a new MSW resonance enhances active-to-sterile neutrino oscillations near the \iso{Cr}{51} emission lines. The solid purple line, for which we have used a baseline of $L = \SI{50}{cm}$, corresponds to the gallium experiments, while the dashed purple line shows the solar neutrino survival probability.  We use the benchmark parameter point from \cref{eq:benchmark}, which explains the gallium anomaly while being consistent with all constraints. We show in cyan and orange the monochromatic neutrino energies from \iso{Cr}{51} and \iso{Ar}{37} sources. The BOREXINO measurements of the \iso{Be}{7} and \textit{pp} neutrino fluxes, including their $1\sigma$ uncertainties, are shown as red data points with error bars \cite{BOREXINO:2018ohr}. The blue and red arrows at the bottom of the plot indicate the energy ranges of  CNO and $pp$ solar neutrinos, respectively.  The survival probability range preferred by the gallium anomaly \cite{Giunti:2022btk} is shown in green.}
	\label{fig:dark msw}
\end{figure}

With the above constraints in mind, we have constructed a benchmark point which satisfies all of them. The benchmark point features an MSW resonance at $E_\nu^\text{res}\approx \SI{750}{keV}$ and an oscillation length $L \lesssim \SI{1}{m}$ (smaller than the baseline of gallium experiments) for active-to-sterile transitions at this energy. It thus successfully explains the gallium anomaly. The benchmark point reads
\begin{align}
  \theta_{e4}^{\text{vac}} = 0.2^\circ , \qquad
  m_s^2 \approx \Delta m^2 \equiv \SI{100}{eV^2} \,,
  \label{eq:benchmark}
\end{align}
where $m_s$ is the mass of the mostly sterile mass eigenstate.  The matter potential at the benchmark point is $V = \SI{6.7e-5}{eV}$, which corresponds to $g_s/m_\phi = \SI{0.311}{eV^{-1}}$. In \cref{fig:dark msw}, we show in purple the electron neutrino survival probability as a function of $E_\nu$ for this benchmark scenario. It reads
\begin{align}
  P(\nu_e\to\nu_e) = 1 - \sin^2 2 \theta_{e4}^\text{eff}
                          \sin^2\left(\frac{\Delta m^2_{\text{eff}} L}{4 E_\nu}\right) \,,  
  \label{eq:2f}                          
\end{align}
where $\Delta m^2_{\text{eff}} = \Delta m^2 \, (\sin 2 \theta_{e4}^\text{vac}/\sin 2 \theta_{e4}^\text{eff})$. (We have checked that electron neutrino survival probability in the 2-flavor picture matches the one in the full 6-flavor picture; in other words, \cref{eq:2f} suffices here.)

For comparison, in \cref{fig:dark msw}, we also show in green the region in which the gallium anomaly can be explained; to draw this band we have adopted the union of the allowed $3\sigma$ regions given different models for the cross section to excited states of \iso{Ge}{71} \cite{Giunti:2022btk}. In dashed we also show the survival probability of solar electron neutrinos, computed using the adiabatic approximation.
The energies of neutrinos from the $\iso{Cr}{51}$ and \iso{Ar}{37} sources used in gallium experiments are indicated by vertical lines, and those of solar neutrinos are shown as well. As the resonance is narrow, we choose to center it around the energy of the dominant neutrino line from the \iso{Cr}{51} source at $\sim 750$ keV. As can be seen from the figure, this implies that we cannot simultaneously explain the anomalous results for the \iso{Ar}{37} source; this does not seriously limit the model's ability to explain the gallium anomaly as the deficit observed with the \iso{Ar}{37} source is statistically not very significant \cite{Abdurashitov:2005tb}. 

Let us finally comment on early Universe observables, which in general tightly constrain eV-scale sterile neutrinos with large active--sterile mixing. Considering only vacuum mixing, our scenario is safe from these constraints because sterile neutrinos will never thermalize thanks to their small mixing angle \cite{Hannestad:2012ky}.\footnote{Note that the fact that our sterile neutrino interacts with DM does not change this conclusion, unlike in other scenarios in which sterile neutrinos possess ``secret interactions''\cite{Hannestad:2013ana, Dasgupta:2013zpn, Saviano:2014esa, Chu:2015ipa, Forastieri:2017oma, Chu:2018gxk, Forastieri:2019cuf}. In these scenarios, frequent hard scatterings enhance the production of sterile neutrinos, whereas in ours, the tiny coupling between neutrinos and DM, $g_s \approx \num{0.311e-24} \times (m_\phi/\SI{e-24}{eV})$, such scattering is negligible.}  The presence of the MSW resonance, however, may change this picture. To see if the active-sterile mixing angle in the early Universe is resonantly enhanced we calculate the redshift-dependence of the MSW potential and compare it to the neutrino energy. In the early Universe, the DM density is increased according to $\rho(z) = \rho_0 (1+z)^3$, where $z$ is the redshift and $\rho_0$ is the fuzzy DM density today. The increased fuzzy DM density shifts the resonance to lower energies such that at the time of Big Bang Nucleosynthesis (BBN) -- the earliest epoch that can be constrained observationally -- the resonance is at $E_\nu^\text{res}(\text{BBN}) \sim \SI{e-9}{eV}$, which is much smaller than typical neutrino energies at this redshift, $E_\nu \approx \mathcal{O}(\si{MeV})$. Therefore, active--sterile mixing is suppressed at BBN. However, at a redshift of $z \sim \num{7000}$, the average neutrino energy matches the resonance energy. Neutrinos are already decoupled from the plasma at this point, so the resonance leads to the production of sterile neutrinos via adiabatic conversion of active ones, without affecting the total energy density in the neutrino sector. By the time of recombination, sterile neutrinos have turned non-relativistic, implying that analyses of the cosmic microwave background (CMB) would measure an effective number of relativistic neutrino species, $N_\text{eff}$, well below the SM value. Moreover, sterile neutrinos would act as warm DM, which would unfavorably affect structure formation. One possibility of avoiding these constraints is to introduce a new decay channel for the sterile neutrino, with an active neutrino and an auxiliary scalar in the final state.\footnote{Due to the small coupling between sterile neutrinos and DM of $g_s \approx \num{0.331e-24} \times (m_\phi/\SI{e-24}{eV})$, decays involving the DM particle would be too slow.} This new scalar would need to be light enough to act as radiation at least until recombination, or it would need to quickly decay into fuzzy DM or active neutrinos before the temperature of the Universe drops to $T \approx \SI{100}{eV}$ to avoid CMB constraints.

The small active--sterile mixing angle at our benchmark point also suppresses any potential effect of neutrino--DM interactions on neutrino free streaming -- the cross section for such scatterings is proportional to $g_s^4$. 

In summary, the scenario presented here appears to successfully explain the gallium anomaly, while being consistent with all terrestrial and, with a small extension of the model, also with cosmological probes. Nevertheless, its success comes at the expense of moderate fine-tuning between the MSW resonance energy and the energy of the dominant \iso{Cr}{51} emission line. In the future, our scenario could be probed by making use of CNO neutrinos (which have been recently observed for the first time \cite{BOREXINO:2020aww, BOREXINO:2022abl}), given that their energies extend over the whole range of interest for the required resonance. In addition, a precise measurement of neutrinos from a \iso{Zn}{65} source at BEST \cite{Gavrin:2019rtr} would be illuminating. Our prediction is that \emph{no} neutrino deficit should be observed with such a source as the neutrino energy of \SI{1.35}{MeV} is well above the MSW resonance energy. The same would be true for an experiment with an \iso{Ar}{37} source.

\subsubsection{MSW resonance from Interaction with Dark Energy}
\label{subsec:DE}

The scenario introduced above, namely an interaction between sterile neutrinos and fuzzy DM, is not the only option for realizing an MSW resonance aligned with the \iso{Cr}{51} emission lines. An alternative is a coupling between sterile neutrinos and dark energy in the form of a vector field \cite{Boehmer:2007qa, Tasinato:2014eka, Tasinato:2014mia}. The equations of motion from Ref.~\cite{Tasinato:2014eka} show that the value of the dark energy field today would need to be $\phi \gtrsim \SI{e7}{eV}$. Looking at our \cref{eq:mswdark}, the MSW potential reads approximately $\phi g_s$, so our benchmark point with $V = 6.7\times 10^{-5}$ eV would correspond to $g_s \lesssim 10^{-11}$. With such a tiny coupling in the dark sector and in the absence of kinetic mixing between the SM photon field and the field representing the dark energy, the presented scenario is experimentally viable. As in \cref{subsec:MSW}, cosmological constraints again require that the field value, $\phi$, was larger in the early Universe than it is today to avoid abundant production of sterile neutrinos before BBN.  Production after BBN may then still be a problem, but could be resolved in the same way as in \cref{subsec:MSW} by introducing a new decay mode for $\nu_s$.

\subsubsection{Parametric resonance}
\label{subsec:parametric}

Another avenue for enhancing oscillations between active and sterile neutrinos is through a parametric resonance \cite{Petcov:1998su, Akhmedov:1998ui}. Such resonances have recently been discussed in the context of neutrino couplings with ultralight DM in Ref.~\cite{Losada:2022uvr}. Unlike in \cref{subsec:MSW}, we assume here that DM is not a polarized vector boson, but either an ultralight scalar, $\phi$, or an unpolarized vector field, $\phi^\mu$.  As far as the phenomenology of neutrino oscillations is concerned, the two cases are equivalent \cite{Brdar:2017kbt} and we will derive all expressions assuming a scalar field. In the neutrino mass basis, the Yukawa interaction between the active ($\nu$) and the sterile ($\nu_4$) state is off-diagonal with a coupling strength $y$:
\begin{align}
    \mathcal{L} \supset y \, \phi \overline{\nu}  \nu_4
    \qquad\text{or} \qquad
    \mathcal{L} \supset y \, \phi^\mu \overline{\nu}  \gamma_\mu \nu_4  \,.
    \label{eq:lagS}
\end{align}
Here, we work for simplicity in the 2-flavor approximation, see below for a discussion of this assumption. \Cref{eq:lagS} effectively modifies the neutrino mass matrix. In particular, it induces flavor transitions between active and sterile neutrino even in the absence of vacuum mixing. This effect is similar to Rabi oscillations in atoms subject to an oscillating electromagnetic field. In our case, the role of the external oscillating field is played by ultralight DM. In such a scenario, the active--sterile mixing can even be taken to vanish completely in the absence of the DM background and only be non-zero when a resonance condition is realized.  This condition reads \cite{Losada:2022uvr}
\begin{align}
    \frac{m_s^2}{4 E_{\nu}} = \frac{m_\phi}{2} \,.
    \label{eq:param-res-condition}
\end{align}
In other words, the oscillation frequency, $m_s^2/(4 E_{\nu})$, should equal half the oscillation frequency of the scalar field, $m_\phi/2$. In this case the quantity
\begin{align}
    \epsilon_\phi \equiv \frac{2 y}{m_s} \frac{\sqrt{2\rho_\phi}}{m_\phi}
    \label{eq:epsilon-phi}
\end{align}
controls the height of the resonance peak, i.e.\ the maximum transition probability. In the two-flavor approximation, and assuming zero vacuum mixing, the latter is given by \cite{Losada:2022uvr}
\begin{align}
  P_{\alpha\beta}^\text{res} = \sin^2 \bigg(\frac{\epsilon_\phi m_\phi L}{4} \bigg) \,,
  \label{eq:res_par}
\end{align}
where $L$ for gallium experiments is around \SI{50}{cm}.  This expression follows from the full transition probability (without vacuum mixing) \cite{Losada:2022uvr}
\begin{align}
    P_{\alpha\beta} = \frac{\epsilon_\phi^2 (1 - \delta_E)^2}{\epsilon_\phi^2 (1 - \delta_E)^2 + 4\delta_E^2}
                      \sin^2 \Big(\frac{m_\phi L}{4} \sqrt{\epsilon_\phi^2 (1 - \delta_E)^2 + 4\delta_E^2} \Big)
    \label{eq:full}
\end{align}
in the limit $\delta_E \equiv 1-m_s^2/(2 E_\nu m_\phi)\to 0$.

\begin{figure}[t]
	\centering
	\includegraphics[scale=0.7]{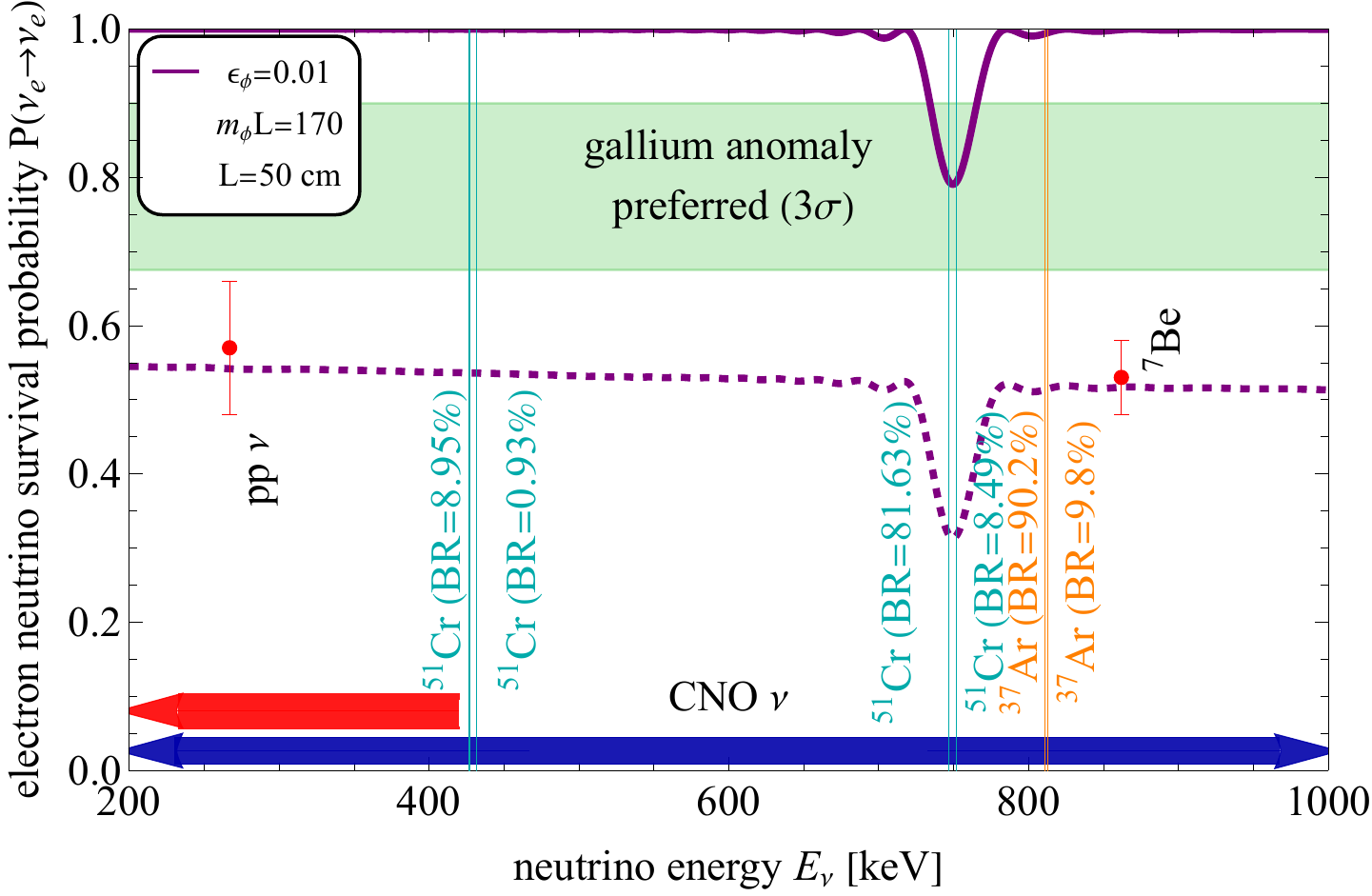} 
	\caption{The electron neutrino survival probability as a function of energy for the scenario from \cref{subsec:parametric}, where a parametric resonance enhances active-to-sterile neutrino oscillations near the \iso{Cr}{51} emission lines.  The solid purple line corresponds to the gallium experiments ($L = \SI{50}{cm}$), while the dashed purple line shows the solar neutrino survival probability. We use the benchmark parameter point from \cref{eq:benchmark-param-res}, which explains the gallium anomaly while being consistent with all constraints. We show in cyan and orange the monochromatic neutrino energies from \iso{Cr}{51} and \iso{Ar}{37} sources. The BOREXINO measurements of the \iso{Be}{7} and \textit{pp} neutrino flux, including their $1\sigma$ uncertainties, are shown as red data points with error bars \cite{BOREXINO:2018ohr}. The blue and red arrows at the bottom of the plot indicate the energy ranges of  CNO and $pp$ solar neutrinos, respectively. The survival probability range  preferred by the gallium anomaly \cite{Giunti:2022btk} is shown in green.}
	\label{fig:parametric res}
\end{figure}

As in \cref{subsec:MSW}, we choose the resonance energy such that it coincides with the dominant \iso{Cr}{51} lines at $E_\nu^\text{res} \approx 750$ keV. Roughly, we need 20\% neutrino disappearance to explain the gallium anomaly, and \cref{eq:res_par} then suggests that $\epsilon_\phi m_\phi L\sim 2$. At the same time, we also require the width of the resonance (which is  $\Delta E_\nu^\text{res}=E_\nu^\text{res}\epsilon_\phi$ \cite{Losada:2022uvr}) to be $\lesssim \SI{10}{keV}$ in order not to induce significant disappearance of reactor and solar neutrinos.  A benchmark point where these requirements are satisfied is given by
\begin{align}
    \epsilon_\phi = 0.01 \,, \qquad m_\phi L = 170 \,.
    \label{eq:benchmark-param-res}
\end{align}
For this benchmark point, $m_s \approx \SI{10.0}{eV}$ (see \cref{eq:param-res-condition}), $y \approx 0.0016$ (see \cref{eq:epsilon-phi}) and $m_\phi\approx \SI{6.7e-5}{eV}$.  In \cref{fig:parametric res} we show the electron neutrino survival probability (one minus the transition probability from \cref{eq:full}) at this benchmark point.

To discuss terrestrial and solar neutrinos, we will once again need to go beyond the 2-flavor approximation. As in \cref{subsec:MSW}, we extend the model to include not one, but three, sterile neutrinos to avoid unacceptable modifications to the active neutrino mixing angles. If we do so, the DM-induced time-dependent active--sterile mixing angle at our benchmark point is never larger than $0.57^\circ$, which implies that, outside the narrow parametric resonance region, oscillation probabilities remain the same as in the SM.

We note that the neutrino--DM interaction which we introduce here induces invisible decays of $Z$ bosons and mesons involving the ultralight DM field, which are constrained \cite{Blinov:2019gcj, Brdar:2020nbj}. We have checked that our benchmark point satisfies these limits. In addition, the parametric resonance scenario faces cosmological constraints.  Notably, the Yukawa interaction from \cref{eq:lagS} with the sizeable Yukawa coupling at our benchmark point could lead to efficient production of $\nu_4$, in violation of constraints on the effective number of relativistic degrees of freedom, $N_\text{eff}$. To avoid such constraints, one could invoke the dynamics of $\phi$ in the early Universe: assuming that the dark matter we observe today is produced via the misalignment mechanism \cite{Preskill:1982cy, Abbott:1982af, Dine:1982ah}, we can postulate that, at BBN, $\phi$ was still rolling down its potential. Its field value during the BBN epoch could then have been much larger than today, rendering $\nu_4$ very heavy through a coupling of the form $\phi \, \bar\nu_4 \nu_4$ \cite{Farzan:2019yvo}.  After $\phi$ has settled down at its present-day value, sterile neutrinos could still be produced, but only at the expense of the energy density stored in the by-then decoupled active neutrinos.  After the temperature has dropped below $m_s \approx \SI{10}{eV}$, sterile neutrinos will quickly decay back to active neutrinos and relativistic $\phi$ particles, so that $N_\text{eff}$ at recombination will again be close to its SM value.  Let us also point out that the value of the Yukawa coupling $y$ at our benchmark point is sufficiently large ($y \gg 10^{-5}$) to prevent neutrinos and DM particles from leaving a hot proto-neutron star unhindered. They will remain trapped and therefore will not lead to significant anomalous cooling, meaning that limits from supernovae are evaded.

A particular concern is the propagation of astrophysical neutrinos.  Notably, scattering of ultra-high energy neutrinos from distant cosmic ray sources on the relic neutrino background could significantly reduce the neutrinos' optical depth.  Since neutrino point sources at cosmological distances have been observed \cite{IceCube:2018cha,IceCube:2022der}, this is disfavored.  At our benchmark point, this could be a problem if the dark matter is scalar, but not if it is a vector field, given that the cross section for the process $\nu\nu \to \nu_4 \nu_4$ in the latter case is significantly smaller \cite{Brdar:2017kbt}.

\subsection{A Decaying Sterile Neutrino}
\label{sec:bsm-nu-s-decay}

Another interesting scenario, which has already been studied to explain the LSND \cite{LSND} and MiniBooNE anomalies \cite{MiniBooNE:2008hfu, MiniBooNE:2008yuf, MiniBooNE:2018esg}, is a decaying sterile neutrino. The idea there is that the sterile neutrino gets produced via mixing with muon neutrinos and then subsequently sources active neutrinos (mainly electron neutrinos) via its decay \cite{Palomares-Ruiz:2005zbh, Dentler:2019dhz, deGouvea:2019qre, Brdar:2020tle,Hostert:2020oui}.

The idea in the context of gallium anomaly would be to invoke a very short-lived sterile neutrino state which decays just outside of the detector in gallium experiments. This way, gallium experiments can experience a sizeable deficit of active neutrinos, but the ``missing'' flux is quickly regenerated, alleviating the otherwise strong constraints from reactor and solar experiments. The baselines of gallium experiments are $\sim \SI{50}{cm}$ in BEST \cite{Barinov:2022wfh}, $\sim \SI{70}{cm}$ in SAGE \cite{Abdurashitov:1998ne}, or \SI{1.90}{m} in GALLEX \cite{Anselmann:1994ar}, while the reactor experiments with the shortest baselines are located $\gtrsim \SI{8}{m}$ from the reactor core \cite{Kwon:1981ua, Alekseev:2016llm, PROSPECT:2015iqr, STEREO:2018blj}.

The  maximal neutrino energy achievable in \iso{Cr}{51} (or \iso{Ar}{37}) decay presents a kinematic upper limit on the sterile neutrino mass of around \SI{750}{keV}.%
\footnote{Heavier sterile neutrinos, which are kinematically inaccessible, would lead to a non-unitary mixing matrix \cite{Antusch:2006vwa, Denton:2021mso}. To explain the gallium anomaly in this way, the normalization of the first (electron flavor) row of the leptonic mixing matrix would need to different from unity by 15--20\%, but has been constrained to be smaller than $\mathcal{O}(1\%)$ \cite{Parke:2015goa, Ellis:2020hus, Ellis:2020ehi, Hu:2020oba}.}
Sterile neutrinos are also severely constrained by searches for kinks in beta decay spectra \cite{Atre:2009rg, Bolton:2019pcu, KATRIN:2022ith, Giunti:2022btk, Abdullahi:2022jlv}; to avoid these limits, we need to resort to sterile neutrino masses of $\lesssim \SI{10}{eV}$. In this mass range, the decay rate of the sterile neutrino into three active neutrinos via Z boson exchange is too small to allow for the short decay lengths $< \SI{10}{m}$ that we are interested in. Therefore, we need to introduce a new decay channel to allow for faster decays. 

To do so, we extend the model with a very light scalar mediator, $S$, which couples to sterile neutrinos, $\nu_s$, via a Yukawa interaction of the form $y_s \, S \overline \nu_s \nu_s$. Via mixing, the sterile neutrino can then decay to $S + \nu_e$, with a decay rate $\propto (y_s^2 m_s/16\pi^2) \sin^2 \theta_{e4}$. For simplicity, we have here assumed that $\nu_s$ mix only with electron neutrinos.  We assume that, unlike in the model introduced in Ref.~\cite{Dentler:2019dhz}, $S$ cannot further decay into active neutrinos due to its small (or even vanishing) mass. Therefore, the constraints from the non-observation of solar antineutrinos do not apply \cite{Hostert:2020oui}.  For $y_s \lesssim 1$, it is now possible to achieve sterile neutrino decay lengths as short as \SI{1.9}{m} to \SI{4}{m}. The fraction of $\nu_s$ that survive out to $\sim \SI{10}{m}$ is then already $< 0.1$. The total flux of active neutrinos passing through a typical detector at a reactor site would hence be similar to the flux in the SM.

However, there are two effects which nevertheless put the model in tension with reactor data. Firstly, the energy of active neutrinos from sterile neutrino decay is smaller by approximately a factor of two than the energy of the primary $\nu_s$, and since the detection cross section scales with $E_\nu^2$, this implies a suppression in the measured event rate. Secondly, around 80\% of the reactor antineutrino flux are at neutrino energies below $\sim \SI{3.5}{MeV}$; for sterile neutrinos produced in this energy range, the regenerated active neutrinos would typically have energies below the threshold for detection via inverse beta decay ($\sim \SI{1.8}{MeV}$).  In the end, the regeneration effect is therefore only at the level of 1\% (relative difference between event rates with and without regeneration for a reactor neutrino detector at a distance of \SI{10}{m}). Clearly, this does not resolve the tension between gallium and reactor data. By similar arguments, also the tension with solar data is only mildly alleviated. On the other hand, the decaying sterile neutrino scenario would engender a mild improvement in the global fit when compared to the vanilla eV-scale sterile neutrino scenario.

Unlike the model from Refs.~\cite{Palomares-Ruiz:2005zbh, Dentler:2019dhz, deGouvea:2019qre, Hostert:2020oui}, the scenario discussed here is less constrained by perturbativity arguments thanks to the lower neutrino energies which allow fairly short decay lengths even for Yukawa couplings $\lesssim 1$.  On the other hand, we require lighter sterile neutrinos to accommodate $\mathcal{O}(\text{10--20\%})$ mixing in view of constraints from beta decay spectra. The large mixing implies that active neutrinos feel the new interaction mediated by $S$ with only mild suppression, so they start free-streaming relatively late in cosmological history, in conflict with CMB observations. Additional new physics would be required to resolve this problem \cite{Chu:2018gxk}.  Supernova constraints, on the other hand, are evaded for the same reason as in \cref{subsec:parametric}: the new interaction is so strong that neutrinos and $S$ particles cannot free-stream out of a supernova core.

\section{Conclusions}
\label{sec:conclusions}

For the past decade, several anomalies have kept the neutrino community on their toes.  But with the reactor anomaly fading away and reactor experiments consequently shifting their focus to other opportunities \cite{Giunti:2021kab, Berryman:2021yan, Letourneau:2022kfs, STEREO:2022nzk} and the situation surrounding the MiniBooNE anomaly more puzzling than ever \cite{Brdar:2021ysi, MicroBooNE:2021tya, Arguelles:2021meu, Denton:2021czb, Kelly:2022uaa}, the gallium anomaly now takes center stage, with a $\gtrsim 4\sigma$ effect claimed by BEST \cite{Barinov:2021asz} and a $\gtrsim 5\sigma$ effect emerging when all gallium experiments are combined \cite{Giunti:2022xat}. In the present work we have scrutinized several possible ways to explain the gallium anomaly, both within the SM and beyond. All of them are summarized, and rated, in \cref{tab:summary}, where we also list previous attempts from the literature.

As a leading candidate for the explanation of the gallium anomaly within the SM, we regard a possible problem with the measurement of the radiochemical \iso{Ge}{71} extraction efficiency.

Regarding BSM explanations, we have identified several promising scenarios consistent with all constraints, but requiring a fine-tuned MSW or parametric resonance.

Future probes of the proposed solutions will hopefully be able to shine light on the origin of the gallium anomaly.
A probe of the proposed SM solution involving the translation of the heat output of the Cr source to the neutrino flux could come from 
using a different source. A more precise measurement with an Ar source or involving a \iso{Zn}{65} source at BEST \cite{Gavrin:2019rtr} does not require the knowledge of  $\text{BR}(\iso{Cr}{51} \to \iso{V^*}{51})$ and can therefore test a bias in the translation of heat output to neutrino flux.
On the other hand, SM explanations which involve neutrino capture on Ga and extraction of Ge will remain viable explanations if future measurements using different sources confirm the anomaly. Only a real-time measurement (like the ill-fated SOX experiment at the BOREXINO detector \cite{Borexino:2013xxa}) or a radiochemical experiment using a different target isotope (for instance \iso{Cl}{37} in combination with a \iso{Zn}{65} source) could resolve those.

Finally, an interesting complementary probe of the proposed SM solutions consists of using a \iso{Cr}{51} source and a Ga doped scintillation detector as proposed in \cite{Huber:2022osv}.  As in this case, in addition to the neutrino capture signal, also elastic neutrino--electron scattering processes can be recorded, the source strength can be determined from the latter process, and by comparing the event rates an incorrect capture cross section could be excluded. This method also does not rely on the radiochemical germanium extraction efficiency as the measurement is done in real time.

Regarding extensions of the SM, the scenarios involving a resonance were tuned such that the resonant enhancement of the oscillations happens at the energies of the dominant Cr lines. Therefore, such scenarios can be probed using a different source which emits neutrinos with a different energy.  
Our prediction is that \emph{no} neutrino deficit should be observed in such an experiment with, for example, a Zn or Ar source.
Furthermore, the BSM scenario of a sharp resonance could be probed by making use of CNO neutrinos as their energies extend over the whole range of interest for the required resonance. 

In summary, we have put forward new avenues for explaining the gallium anomaly both in the SM and beyond, and we have provided an overview of previously proposed BSM solutions. We encourage new experimental efforts exploring these avenues in the future. A discovery of an explanation beyond the SM would clearly constitute a major revolution in particle physics; but even if a SM explanation for the gallium anomaly is uncovered, we will still have learned important lessons that may guide the design and interpretation of future experiments.

\begin{table}
  \centering
  \caption{A summary of explanation attempts for the gallium anomaly. The first part of the table explores solutions within the SM, the second part contains potential solutions that require new physics. Besides the scenarios discussed in the present paper, we also include proposals from the literature. (We do not include star ratings for the latter.)}
  \vspace{0.2cm}
  \def\arraystretch{1.3}
  \begin{tabular}{p{0.2cm}p{4.6cm}p{9.5cm}c}
    \toprule
    & scenario & comments & our rating \\
    \midrule
    \multicolumn{4}{l}{\bf Explanations within the Standard Model} \\
      & increased \iso{Ge}{71} half-life
          \newline \quad (\cref{sec:Ge71-decay} and Ref.~\cite{Giunti:2022xat})
          & would lead to smaller matrix element for $\nu + \iso{Ga}{71}$; but the \iso{Ge}{71} half-life has been measured many times with different methods in \cite{Hampel:1985zz}, all of which yield consistent results. So it is hard to imagine a bias in these measurements. 
          & \fstar\fstar\ostar\ostar\ostar \\
      & new \iso{Ga}{71} excited state \newline \quad (\cref{sec:Ga71-excitation})
          & would imply a bias in the extraction of the $\nu + \iso{Ga}{71}$ matrix element from the measured $\iso{Ge}{71}$ half-life. Some very old experiments claim the existence of such a state, but this has not been confirmed in more recent observations.
          & \fstar\fstar\ostar\ostar\ostar \\
      & increased $\text{BR}(\iso{Cr}{51} \to \iso{V^*}{51})$ \newline \quad (\cref{sec:Cr51})
          & would cause a bias in translating the heat output of the source to a neutrino production rate. Measurements of $\text{BR}(\iso{Cr}{51} \to \iso{V^*}{51})$ show some tension, but it is far less than the shift required to explain the gallium anomaly.
          & \fstar\fstar\fstar\ostar\ostar \\
      & \iso{Ge}{71} extraction efficiency \newline \quad (\cref{sec:calibration})
          & one of SAGE's calibration runs has revealed a large bias. Could a small, unnoticed, bias have been present in all gallium experiments?
          & \fstar\fstar\fstar\fstar\ostar \\
    \midrule
    \multicolumn{4}{l}{\bf Explanations beyond the Standard Model} \\
      & $\nu_s$ coupled to ultralight DM \newline (MSW resonance, Sec.~\ref{subsec:MSW})
          & several exotic ingredients; somewhat tuned MSW resonance;
          new $\nu_4$ decay channel required for cosmology.
          & \fstar\fstar\fstar\fstar\ostar \\
   
      & $\nu_s$ coupled to dark energy \newline (MSW resonance, Sec.~\ref{subsec:DE})
          & several exotic ingredients; somewhat tuned MSW resonance;
          cosmology similar to the previous scenario.
          & \fstar\fstar\fstar\ostar\ostar \\
   
      & $\nu_s$ coupled to ultralight DM \newline (param. resonance, Sec.~\ref{subsec:parametric})
          & several exotic ingredients; somewhat tuned parametric resonance;
          cosmology requires post-BBN DM production via misalignment.
          & \fstar\fstar\fstar\fstar\ostar \\
   
      & decaying $\nu_s$ \newline \quad (\cref{sec:bsm-nu-s-decay})
          & difficult to reconcile with reactor and solar data; regeneration of active neutrinos in $\nu_s$ decays alleviates tension, but does not resolve it.
          & \fstar\fstar\ostar\ostar\ostar \\
          
      & vanilla eV-scale $\nu_s$ \newline (Refs.~\cite{Barinov:2022wfh, Giunti:2022btk})
          & preferred parameter space is strongly disfavored by solar and reactor data.
          & \fstar\ostar\ostar\ostar\ostar \\

      & $\nu_s$ with CPT violation \newline (Refs.~\cite{Giunti:2010zs})
          & avoids constraints from reactor experiments, but those from solar neutrinos cannot be alleviated.  & \\ 

      & extra dimensions \newline (Refs.~\cite{Machado:2011kt, Carena:2017qhd, Forero:2022skg}) 
          & neutrinos oscillate into sterile Kaluza--Klein modes that propagate in extra dimensions; in tension with reactor data.
          & \\
          
      & stochastic neutrino mixing \newline (Ref.~\cite{Zavanin:2015oia})
          & based on a difference between sterile neutrino mixing angles at production and detection (see also \cite{Babu:2021cxe,Babu:2022non}); fit worse than for vanilla $\nu_s$. 
          & \\
          
      & decoherence \newline (Refs.~\cite{Arguelles:2022bvt, Hardin:2022muu})
          & non-standard source of decoherence needed; known experimental energy resolutions constrain wave packet length, making an explanation by wave packet separation alone challenging.  & \\

      & $\nu_s$ coupled to ultralight scalar
          \newline (Ref.~\cite{Davoudiasl:2023uiq})
          & ultralight scalar coupling to $\nu_s$ and to ordinary matter affects sterile neutrino parameters; can not avoid reactor  constraints \\
    \bottomrule
  \end{tabular}
  \label{tab:summary}
\end{table}

\section*{Acknowledgments}

We are grateful to Ivan Esteban, Carlo Giunti, Wolfgang Hampel, and Gilad Perez for useful discussions.

\bibliographystyle{JHEP}
\bibliography{refs}

\end{document}